\documentclass[11pt]{article}
\usepackage{color,amsfonts,amsmath,amssymb,mathrsfs,verbatim,epsf}

\def\inn{\in}

\def\pqo{\mbox{$ \!\!\!\!\; \mbox{\textbf {\Large \_}} \! $}}
\def\sp{\; \!}

\def\kl{k \pqo l}
\def\jl{j \pqo l}

\def\il{i \mbox{$ \!\!\!\!\; \:\! \mbox{\textbf{\Large \_}} \! \!\!\; $} l}

\def\TM{\mbox{\it TM}}

\def\rrr{{\mathbb{R}}}
\def\ccc{{\mathbb{C}}}
\def\hhh{{\mathbb{H}}}
\def\ooo{{\mathbb{O}}}

\def\b1{\mbox{\boldmath $1$}}
\def\bbb{\mbox{\boldmath $3$}}
\def\v{\mbox{\boldmath $v$}}

\def\bh{\mbox{\boldmath $h$}}

\def\bv{\mbox{\boldmath $v$}}

\def\lvtfep{L_8(\v_{248})_{\mathrm{E}_8}=1}

\def\htwc{\mbox{h}_2\ccc}
\def\hthc{\mbox{h}_3\ccc}

\def\htwo{\mbox{h}_2\ooo}
\def\htho{\mbox{h}_3\ooo}

\def\sltc{\mbox{SL}(2,\ccc)}

\def\slthc{\mbox{SL}(3,\ccc)}

\def\sltho{\mbox{SL}(3,\ooo)}

\def\sltwoo{\mbox{SL}(2,\ooo)}

\def\uo{\mbox{U}(1)}

\def\sutw{\mbox{SU}(2)}

\def\suth{\mbox{SU}(3)}

\def\soot{\mbox{SO}^+(1,3)}

\def\sootnm{\mbox{SO}^+(1,n-1)}

\def\gt{\mbox{G}_2}

\def\ee{\mbox{E}_8}

\def\ese{\mbox{E}_7}

\def\esi{\mbox{E}_6}

\def\SML{\mbox{SU}(3)_c \times \mbox{SU}(2)_L \times \mbox{U}(1)_Y}

\def\mcM{{\mathcal M}}

\def\setb{\setlength{\baselineskip}{0.625\baselineskip}}

\begin{document} 

{\setlength{\baselineskip}{0.625\baselineskip}

\begin{center}

 {\LARGE{\bf Octonions in Particle Physics through  \vspace{5pt} \\ 
             Structures of Generalised Proper Time }}

   \vspace{12pt}


  \mbox {{\Large David J. Jackson}} \\ 
  \vspace{6pt}  
  {david.jackson.th@gmail.com}  \\

  
 \vspace{10pt}
 
 { \large September 2, 2019 }

 \vspace{18pt}

{\bf  Abstract}

\vspace{-3pt} 
 
\end{center}


In considering the nature of the basic mathematical structures appropriate for 
describing the fundamental elements of particle physics a significant role for the 
\mbox{octonions}, as an extension from the complex numbers and uniquely the largest 
division algebra, has occasionally been proposed. Rather than being based initially upon 
the more abstract grounds of mathematical aesthetics, here we describe a unified theory 
motivated conceptually through an elementary generalisation of the expression for a local 
proper time interval, beyond that of 4-dimensional spacetime and also beyond that of the 
local structure of models with extra spatial dimensions, for which the explicit 
mathematical development 
 naturally incorporates the octonion algebra as an essential feature. Properties of matter 
are identified directly through the symmetry breaking structure entailed in the necessary 
extraction of the external 4-dimensional spacetime background and are shown to reproduce a 
series of characteristic structures of the Standard Model of particle physics.
 While already employing octonion-based constructions of the exceptional Lie groups $\esi$ 
and $\ese$ 
 the uncovering of the full 
Standard Model, as well as new physics beyond, is predicted to involve an $\ee$-related 
structure for which the octonion algebra is again anticipated to be of fundamental 
importance.

{





\vspace{-1pt} 
\tableofcontents
\vspace{-1pt} 
}



\section{The Octonions and Particle Physics}
\label{octo1}

 One of the most prominent mysteries in physics concerns the origins of 
the elementary properties of matter on the smallest discernible scale 
accessible to experiment as consolidated in the Standard Model of 
particle physics. We hence begin by briefly reviewing the structures to 
be accounted for. 
The elementary particles are classified according to their 
transformation properties under the external 4-dimensional spacetime 
Lorentz symmetry and internal strong, weak and electromagnetic gauge 
symmetries, as listed in table~\ref{smgen} for the first generation of 
leptons and quarks with the weak gauge component omitted.

\begin{table}[htbp]
\centering
\begin{tabular}{|r|ccccc|c|}
 \hline
    $N$ & Lorentz  & $\times$ & $\suth_c$ & $\times$ & $\uo_Q$ 
	        & state   \\
 \hline  
    4 & Weyl  & & $\b1$  & & 0 & $\nu_L$  \\
	& & & & & &
    \vspace{-18pt} \\ 
	8 & Dirac & & $\b1$  & & 1 & 
{\LARGE$\binom{\mbox{\normalsize$e_L$}}{\mbox{\normalsize$e_R$}}$} \\ 
   & & & & & &
    \vspace{-15pt} \\  	
   24 & Dirac & & $\bbb$ & & $\frac{2}{3}$ &  
{\LARGE$\binom{\mbox{\normalsize$u_L$}}{\mbox{\normalsize$u_R$}}$} \\
      & & & & & &
    \vspace{-15pt} \\
   24 & Dirac & & $\bbb$ & & $\frac{1}{3}$ & 
{\LARGE$\;\!\binom{\mbox{\normalsize$d_L$}}
                    {\mbox{\normalsize$d_R$}}\!\!\:$}  
   	\vspace{-15pt}  \\  & & & & & &
	  \\
   \hline
  \end{tabular}
  \caption{\setb The Lorentz spinor structure and transformations under 
the strong $\suth_c$ and electromagnetic $\uo_Q$ gauge groups (with 
relative charge magnitudes listed in the latter case) for a generation 
of $\nu$ and $e$-lepton and $u$ and $d$-quark states in the Standard 
Model. Also listed is the number $N$ of real components needed to 
describe the respective fields.}
\label{smgen}
\end{table}

  In the case for example of the left-handed neutrino the corresponding 
Weyl spinor $\nu_L \in \ccc^2$ incorporates four real components, with 
a doubling to include the right-handed $R$ components for the Dirac 
spinors and a further trebling to incorporate the triplets $\bbb$ under 
$\suth_c$ for the three `colours' of quark states.

  The complete Standard Model structure includes a full three 
generations of leptons and quarks, that is with two further copies of 
the states in table~\ref{smgen} with higher masses, as well as a full 
internal $\SML$ gauge symmetry. The weak $\sutw_L$ factor acts upon 
doublets of left-handed leptons $\binom{\nu_L}{e_L}$ and quarks 
$\binom{u_L}{d_L}$ and mixes states across the three generations, while 
$Y$ is the hypercharge. 
 The action of the electroweak symmetry $\sutw_L \times \uo_Y$ upon a 
scalar Higgs doublet breaks this symmetry to $\uo_Q$ as the Higgs field 
takes a non-zero vacuum value, with each generation of leptons and 
quarks acquiring the electromagnetic $\uo_Q$ charges listed in 
table~\ref{smgen}. Interactions with the Higgs serve as the `origin of 
mass' for all particle states, as parametrised by Yukawa couplings 
which also describe the mass hierarchy across the three generations
(for a more extensive review of the Standard Model see for 
example~\cite{Teub}, \cite{Unifi}  section~7.2).
 
 Generation mixing in the neutrino sector is required in order to 
describe the empirical observations of solar and atmospheric neutrino 
oscillations (\cite{Teub} chapter~7, \cite{PDG18} lepton summary table 
and section~14). 
In the corresponding models this generally implies the introduction of 
two or three right-handed neutrinos states $\nu_R$ to generate mass 
terms for the $\nu_L$ components, with the neutrinos then having the 
Lorentz spinor structure of Dirac or Majorana particles~(\cite{Drewes} 
section~2), as a minimal required extension beyond the Standard Model 
(for which there are no $\nu_R$ states and the $\nu_L$ states are 
massless and don't mix).
    
   A key feature of the Standard Model is the symmetry and 
transformation properties of particle multiplets as described by group 
theory, and in particular by Lie groups and their
 representations~\cite{Geor}. 
  Given the fundamental level of the matter being investigated the 
question is inevitably raised concerning the \textit{source} of the 
specific structures described in and following table~\ref{smgen}, an 
understanding of which would clearly be preferable to arbitrarily 
imposing these properties by hand to match empirical observations as is 
largely the case for the Standard Model itself. That is, intuitively 
and ideally, it might be expected that a well-motivated, unique and 
unifying mathematical framework should be closely connected with, and 
provide an explanation for, the nature of
  this most elementary level of physical structure.
One such line of investigation has been the unique series of five 
exceptional Lie groups, through $\esi$~\cite{Gur1} and 
$\ese$~\cite{Gur7}  to the largest case of $\ee$~\cite{Bars}, as 
candidate unification symmetries. Another approach has placed the focus 
on the unique sequence of four division algebras, of which the largest 
is the octonions, as a proposed source of particle properties, as we 
review in the following.

  As an extension from the real numbers $\rrr$, complex numbers $\ccc$ 
and quaternions $\hhh$, the octonions $\ooo$ have eight real 
coefficients $a_1,\ldots, a_8$ with an individual element $a \in \ooo$ 
written as (\cite{Unifi} equation~6.6, adopting the notation 
of~\cite{Wang,Man4}):
\begin{equation}
\label{octa}
  a \; = \; a_1 \; + \; a_2\,i \; + \; a_3\,j + 
         \; a_4\,k \; + \; a_5\,{\kl}
   \; + \; a_6\,{\jl}   \; + \; a_7\,{\il} \; + \; a_8\,l
\end{equation} 
  The seven imaginary units $\{i,j,k,\kl,\jl,\il,l\}$, with 
$i^2=j^2=\ldots=\il^2=l^2 = -1$, are mutually anticommuting, with for 
example $ k \sp\jl = - \jl\sp k =\il$, with the full algebra 
composition structure described in (\cite{Unifi} figure~6.1, 
\cite{Man4} figure~2, the notation is motivated by compositions such as 
$i\sp l =\il$). While multiplication within any quaternion subalgebra 
is associative there are also in general many anti-associative 
compositions such as $(i\sp j)l = - i(j \;\! l) = +\kl$.
 The octonions do however form an `alternative' algebra since products 
involving any two elements $a,b \in \ooo$ are associative, with
  $(aa)b = a(ab)$ for example.

 Via octonion conjugation $a \to \bar{a}$, with $a_1 \to a_1$ and $a_h 
\to -a_h$ for $h = 2,\ldots, 8$ in equation~\ref{octa}, the octonion 
norm $\vert a \vert$, with $\vert a \vert^2  = \bar{a}a = a\bar{a} = 
\sum_{h=1}^8 a_h^2$, can be defined and also a unique multiplicative 
inverse $a^{-1} = \frac{\bar{a}}{\vert a \vert^2}$ for any $a\neq 0$,
 with $a^{-1}a = aa^{-1} = 1$,
 making the octonions a `normed division algebra'. While the octonions 
were discovered in the 1840s it was proved in the 1890s that the 
octonions are uniquely the largest normed division 
algebra~(\cite{Baez1}~sections~1 and 1.1); their general properties are 
further reviewed in (\cite{Unifi}~section~6.2, \cite{Baez1}).
 A hint towards their possible application in physics is provided by 
the compatibility of the norm with multiplication, that is with:
\begin{equation}
 \label{normp}
  \vert ab \vert = \vert a \vert \vert b \vert
\end{equation}
 for any $a,b \in \ooo$.
  This relation holds generally for normed division algebras,
  as also for elements of $\rrr$, $\ccc$ and $\hhh$, and it applies in 
the case of $\ooo$ due to the alternative property. The property of  
equation~\ref{normp}, and generalisations from it, is key to the 
ability of the octonions to describe rich symmetry structures, for 
example implying
   norm-preserving transformations
 $b \to b' = ab$ with 
  $\vert b \vert = \vert b' \vert$  for any 
   $a,b \in \ooo$ with $\vert a \vert =1$.  However while describing 
symmetry transformations the octonions do not form a group structure 
 since their composition is non-associative
  (with the associativity of the product of elements being a key axiom 
of group theory), and hence care is needed in forming any bridge to 
particle physics for which Lie group theory is central.

  Nevertheless, given that $\rrr$ and $\ccc$ are profusely employed in 
particle physics theory, as well as in physics more generally, and with 
$\hhh$ composition describing symmetries associated with geometric 
rotations in a Euclidean space with three or four dimensions (see for 
example~\cite{Unifi} discussion of equations~6.11--6.14), it has been 
natural to enquire whether the octonions $\ooo$ might have a 
fundamental role to play in accounting for the physics of elementary 
particles in the geometric arena of 4-dimensional spacetime.
 Further, rather than an indefinite or loose correspondence, such as 
the observation that there are \textit{four} division algebras ($\rrr$, 
$\ccc$, $\hhh$ and $\ooo$) and 
 \textit{four} fundamental forces (gravity and the three Standard Model 
gauge interactions), clearly a more direct and assured link between the
 unique structures of the division algebras and the 
 unique properties of particle physics  
  would be desired as the basis of a theory.
 
  Even before the Standard Model was formed the possibility of 
utilising the octonions for the theoretical description of particle 
properties had been suggested \cite{Pais,Gamb}. The idea was again 
considered as the Standard Model became established from the mid-1970s
 \cite{Gunay73,Gunay,SoLo,Mori2} (a more extensive compilation of the 
earlier circa 1960s work is listed in \cite{SoLo} references~31--41). 
In more recent years the possibility of an intrinsic connection between 
octonions and particle physics has been revived, developed and further 
debated \cite{Dix1, Dix2, Man4, Mori15, Fur1, Fur2, Rowl}.
 
   Typically the above cited approaches  seek an abstract  
correspondence between mathematical structures associated with the 
octonions and patterns of particle properties observed in the 
laboratory, generally motivated by the uniqueness of the division 
algebras. For example while the smallest exceptional Lie group $\gt$ 
describes the automorphism symmetry of the octonion algebra 
composition, leaving the real unit with coefficient $a_1$ in 
equation~\ref{octa} invariant, there are subgroups $\suth \subset \gt$ 
that also stabilise one of the imaginary units, such as with the 
coefficient $a_8$ in equation~\ref{octa} (see for 
example~\cite{Gunay73} section~5, \cite{Man4} section~4).
 While in (\cite{Gunay73} section~7) such an $\ooo$-based symmetry was 
associated with the global $\suth$ \textit{flavour} symmetry of 
\mbox{Gell-Mann's} quark model, incorporating also an octet of pion and 
kaon states, since \cite{Gunay} it has more generally been associated 
with the local $\suth_c$ \textit{colour} symmetry of strong gauge 
interactions in the quark sector.

   This association of the colour gauge group $\suth_c$ of the Standard 
Model, reviewed above for table~\ref{smgen}, with
   the subgroup of the octonion automorphism group  $\suth \subset \gt 
\equiv \mbox{Aut}(\ooo)$ forms the basis for example of the 
  schemes presented in \cite{Mori2,Dix1,Fur1}.
In attempts to connect with further properties of the Standard Model 
the octonion basis might then be extended by incorporating further 
division algebra structures. In~\cite{Mori2} the strong interaction is 
regarded as an `octonionic gauge theory' to be combined with a 
`quaternionic electroweak theory', as revisited in~\cite{Mori15}.
 In~\cite{Dix1} the kernel structure of the Standard Model is 
associated with the properties of the tensor product space $\ccc \times 
\hhh \times \ooo$ acting upon itself from the left.
 Similarly~\cite{Fur1} is based upon the left-action of $\ccc \times 
\ooo$ on itself to describe strong and electromagnetic gauge 
symmetries, with the properties of a further $\ccc \times \hhh$ factor 
associated with the Lorentz spinor structure of particle states in 
4-dimensional spacetime.
 The augmentations include the need to incorporate a full three 
generations of leptons and quarks, as proposed for example 
in~\cite{Dix2} through further factors of $\hhh$ and $\ooo$, and as 
accommodated  in~\cite{Fur2} through structures generated by $\ccc 
\times \ooo$ with the tensor product $\rrr \times \ccc \times \hhh 
\times \ooo$ suggested as a basis for seeking to replicate the full 
Standard Model.

 While a degree of correspondence with the Standard Model has been 
achieved there is a danger that attempts to attain a more complete 
match may begin to appear somewhat contrived, losing an element of the 
spirit of uniqueness that helped initiate these studies. Further, a 
more robust argument for motivating the connection between octonions 
and particle physics at a fundamental level is still lacking, with for 
example~\cite{Fur2} concluding with the query: ``If the Standard 
Model's group representation structure is indeed a result of the 
algebras $\rrr$, $\ccc$, $\hhh$, and $\ooo$, then what is it exactly 
that is so special about these algebras?''
    
   As described above for equation~\ref{normp} although the octonions 
are non-associative the octonion algebra composition can still be 
utilised to describe symmetry transformations. If such a symmetry can 
be \textit{connected with a basic aspect of the physical world} then 
the non-associativity is not necessarily prohibitive, taking such a 
\textit{physical} symmetry motivated through a clear conceptual 
foundation to have priority over the direct application of 
\textit{mathematical} group theory and its associated \mbox{axiomatic 
formulation}.
 Indeed this was one of the earliest motivations for considering the 
octonions in particle physics. Based on equation~\ref{normp}
 in (\cite{Pais} referring to equations~5 and 6 therein)
 Pais described the possibility of defining `octonion gauge 
transformations' with the non-associative octonions considered to form 
a `quasi-group', while in (\cite{Gamb} opening three paragraphs) Gamba 
similarly argues that fundamental symmetries for particle physics need 
not form a group structure.

 Without direct reliance on a group structure care is needed to 
ascertain at which points the extensive tools of group theory can be 
employed and at which points explicit calculation is required owing to 
the non-associativity. On the other hand with symmetry structures being 
central to the Standard Model, symmetries associated with the octonion 
composition seem a reasonable place to seek a fundamental connection 
with particle physics. In order to construct a physical theory there 
remains the question concerning \textit{how} such symmetries might 
enter physics, that is regarding
 \textit{what kind of physical entity} the octonions might represent or 
act upon in order for these algebraic structures to underlie the 
properties of empirical phenomena.

  The most basic entities in physics could be considered to be our 
conceptions of \textit{space}, \textit{time} and \textit{matter}, with 
the equations of physics describing the properties of and relations 
between these fundamental elements in any theory. The key question then 
concerns \textit{why} the octonions should have anything to do with 
\textit{space}, \textit{time} or \textit{matter} (indeed, 
philosophically, the question can be raised concerning why \textit{any} 
mathematical object should be connected with these or any other 
physical concept). Of these three basic entities for the present theory 
the primary status is assigned to  \textit{time}. 

  In the case of time an argument for a direct and unambiguous  
connection with a mathematical structure can be observed immediately on 
noting that the simple one-dimensional continuous flow of time can be, 
and generally is,  associated with and parametrised by the ordered 
continuum of real numbers $\rrr$, the smallest division algebra. The 
nature of the continuum allows arbitrarily small intervals of time 
$\delta s \in \rrr$ to be considered and, as we shall describe in the 
following 
section, a direct arithmetic analysis and generalisation  will lead to 
a substructure and symmetry properties for the proper time interval 
$\delta s$ with a pivotal role for the composition structure of the 
octonions $\ooo$, the largest division algebra.

   As we also describe in the following section the underlying 
conception of the theory can also be interpreted as a simple 
generalisation from approaches that aim to deduce the properties of 
matter fields from a structure of extra spatial dimensions over 
4-dimensional spacetime. 
 Here the local geometry of 4-dimensional spacetime will be identified 
with a substructure of the \textit{generalised expression for proper 
time}, rather than with the substructure of a 
\textit{higher-dimensional spacetime}, while in both cases the 
properties of the residual components can be associated with matter 
fields in the \mbox{4-dimensional} spacetime background. 
 Since the octonions are intimately involved in this generalisation of 
the local structure of proper time, and are hence
 motivated and rooted at the most elementary conceptual level,
 and since the resulting matter fields are found to resemble 
significant features of the Standard Model, as will be reviewed in 
section~\ref{octo3}, this theory provides a direct means and 
\textit{explanation} for the application of the octonions in particle 
physics.

\section{Generalised Proper Time and Extra Dimensions}
\label{octo2}

 In pursuing the ambition of accounting for the empirical properties of 
matter as observed in 4-dimensional spacetime one major avenue of 
approach has been to augment 4-dimensional spacetime itself with extra 
spatial dimensions (see for example~\cite{Rizzo,Liu} and references 
therein). On applying a kind of `higher-dimensional origami' to fold up 
or `compactify' the extra components the aim is then to deduce a 
pattern of properties that would ideally resemble the structure of 
physical matter fields. While providing a unifying geometrical 
framework, as initiated by Kaluza and Klein with the addition of a 
single extra dimension proposed to accommodate 
electromagnetism~\cite{Kaluza,Klein}, extensions with further spatial 
dimensions in more modern theories have only achieved limited success 
in accounting for the Standard Model~\cite{Witt}. Indeed it has proved 
difficult even to accommodate the Standard Model within a suitably 
contrived structure of extra spatial dimensions (see for 
example~\cite{Jitt}), with the goal of finding a unique and inevitable 
explanation for the properties of particle physics in this manner 
seemingly still somewhat remote.

  Since we are interested in the most elementary properties of matter  
on the smallest discernible scale here we zoom in and first consider 
the geometry of an infinitesimal local inertial reference frame in 
4-dimensional spacetime within which an infinitesimal proper time 
interval $\delta s$ can be expressed as:
\begin{equation}
  \label{sfourdo} 
   (\delta s)^2   \; = \; \eta_{ab}\delta x^a \delta x^b  
   \quad \mbox{for} \quad a,b = 0,1,2,3
\end{equation}  
  with $\eta = \mbox{diag}(+1,-1,-1,-1)$ the  Lorentz metric
  and $\{x^a\}$ a set of local inertial frame coordinates (the 
summation convention for repeated indices is assumed throughout this 
paper). The local coordinates are arbitrary within local Lorentz 
$\soot$ transformations leaving $\delta s$ invariant.

  While assumptions concerning the extended higher-dimensional 
\textit{global} geometric structure vary from model to model, 
essentially all frameworks with extra spatial dimensions by definition 
\textit{locally} augment equation~\ref{sfourdo} directly to the form:
\begin{equation}
  \label{sndo} 
   (\delta s)^2   \; = \; \hat{\eta}_{ab}\delta x^a \delta x^b  
   \quad \mbox{for} \quad a,b = 0,\ldots,n-1
\end{equation}     
 with $n>4$ and with $\hat{\eta} = \mbox{diag}(+1,-1, \ldots, -1)$ the 
augmented local Lorentz metric structure for the $n$-dimensional 
spacetime, now with a local $\sootnm$ symmetry.

  However, since we do not perceive or navigate around the extra 
dimensions and hence do not require them to possess the quadratic form 
of any Euclidean geometric properties,
 and since here we are exploring very local structure in the limit 
$\delta s \to 0$, for the present theory we note that 
equation~\ref{sndo} can be further generalised to the 
$p^{\mathrm{th}}$-order homogeneous polynomial form:  
\begin{equation}
 \label{salphao}
  (\delta s)^p  \; = \; \alpha_{abc\ldots}\delta x^a 
                            \delta x^b \delta x^c \ldots
    \quad \mbox{for} \quad a,b,c,\ldots = 0,\ldots,n-1
\end{equation}
  with $n>4$ \textit{and} also $p>2$, with each coefficient
$\alpha_{abc\ldots} \inn \{-1,0,1\}$ generalising the components of the 
Lorentz metric, and with the Lorentz symmetry further generalised to a 
full symmetry denoted $\hat{G}$. It will of course still be required to 
incorporate and project out the local geometric structure of the 
external 4-dimensional spacetime, and hence the generalisation of 
equation~\ref{salphao} is permitted provided we can identify a 
substructure corresponding to the quadratic form on the right-hand side 
of equation~\ref{sfourdo}, that is equation~\ref{salphao} must be of 
the form:
\begin{equation}
 \label{sfourpo}
  (\delta s)^p  \; = \; \left[
    \eta_{ab}\delta x^a \delta x^b \right]  
	    (\delta x^4, \ldots ,\delta x^{n-1})^{p-2}
   \; + \;  (\delta x^0, \ldots ,\delta x^{n-1})^{p} 
\end{equation}
 Here $a,b = 0,1,2,3$ correspond to external spacetime indices in the 
first term, in which the second factor represents a 
\mbox{$(p-2)^{\mathrm{th}}$-order} polynomial in the remaining $(n-4)$ 
components, while the second term comprises the further 
$p^{\mathrm{th}}$-order polynomial parts of equation~\ref{salphao}. 

 While dropping the seemingly unnecessary assumption that the nature of 
the extra components should be limited to the locally Euclidean 
quadratic form of the three external spatial dimensions, as restricted 
in equation~\ref{sndo}, the form of equation~\ref{salphao} also offers 
a simpler and more conservative foundation for a theory. Here analysis 
of
 equation~\ref{salphao} is
  based simply upon a study of arithmetic expressions for infinitesimal 
intervals of time, that is $\delta s$ on the left-hand side, with the 
passage of time something that we are intimately familiar with, rather 
than upon the hypothesis of extra spatial dimensions, with ${\delta 
x^4, \ldots, \delta x^{n-1}}$ terms appended to the right-hand side of 
equation~\ref{sfourdo} in equation~\ref{sndo}, a structure that we do 
not perceive at all.

  Before considering possible explicit expressions for 
equation~\ref{salphao} we rewrite this equation in a more convenient 
form on dividing both sides by $(\delta s)^p$ and defining the norm:
\begin{equation}
  \label{lpvno}
  L_p(\bv_n)_{\hat{G}} 
  \; := \; 
 \alpha_{abc\ldots} \frac{\delta x^a \delta x^b \delta x^c \ldots} 
                           {(\delta s)^p\;\;}
						  \Big\vert_{\delta s \to 0}   
	\;  = \; 	
    \alpha_{abc\ldots}v^a v^b v^c \ldots \; = \; 1
\end{equation}
 with each $v^a = \frac{\delta x^a}{\delta s} 
          {\big{\vert}}_{\mbox {\tiny $\delta s \! \to \! 0$}}$
		  (\cite{Gener} section~2 equation~13).
 Here $p$ is the homogeneous polynomial power, $n$ is the number of 
components and $\hat{G}$ is the continuous symmetry acting on the 
components $\{v^a\}$ of the $n$-vector $\bv_n \in \rrr^n$ while 
preserving the equality with unity (see also~\cite{Unifi} section~2.1 
equation~2.9).	In augmenting the whole of the expression in 
equation~\ref{sfourdo} rather than the \textit{spatial} aspect alone, 
equation~\ref{lpvno} (as well as its equivalent in 
equation~\ref{salphao}) can be considered a `generalised form of 
\textit{proper time}', with the adjective `proper' denoting invariance 
under symmetry transformations.

  Here we begin then with a clear conceptual picture
  that leads to equation~\ref{lpvno}
   and \textit{then} seek explicit mathematical forms for this 
expression with corresponding symmetries that preserve the norm. In 
particular the notion of \textit{symmetry} takes precedence over the 
constructions of \textit{group theory} specifically. Hence 
norm-preserving compositions utilising the octonions, as exemplified by 
equation~\ref{normp}, can be considered even though the octonions are 
non-associative and do not directly form a group. This is the key 
observation through which the octonions, as the largest division 
algebra and with a rich structure of symmetry properties, enter the 
present theory. In the explicit mathematical development for the 
general form of proper time of equation~\ref{lpvno}, as an augmentation 
from a local \mbox{4-dimensional} spacetime structure, the octonions 
are indeed found to play an essential role in a unique sequence of 
`Russian doll'-like embeddings as summarised in table~\ref{vallo}.    


\begin{table}[htbp]
\centering
\begin{tabular}{|l|rr|l|}
 \hline
   form of time:   & vector space:$\;$ &     & symmetry:    \\
 \hline  
$L_2(\bv_4)_{\mathrm{SL}(2,\ccc)}=1$ & 
   $\bv_4 \equiv \bh \inn \htwc$ & 4-dimensional
    & $\sltc$ $\,\,\;$on quadratic form   
 	   \\ 
$L_3(\bv_9)_{\mathrm{SL}(3,\ccc)}=1$ & 
   $\bv_9 \inn \hthc$ & 9-dimensional
    & $\slthc$ $\;\quad\:\quad$on cubic form   
 	   \\ 
$L_3(\bv_{27})_{\mathrm{E}_6}=1$ & 
   $\bv_{27} \inn \htho$ & 27-dimensional
    & $\esi \equiv \sltho$ on cubic form   
 	   \\ 
$L_4(\bv_{56})_{\mathrm{E}_7}=1$ & 
   $\bv_{56} \inn F(\htho)$ & 56-dimensional
    & $\ese$ $\;\;\;\;\quad\;\;\;\:\quad$on quartic form   
 	   \\ \hline 			   
  \end{tabular}
  \caption{\setb Progression in explicit expressions for the 
generalised form of proper time of equation~\ref{lpvno} subsuming the 
4-dimensional spacetime form of equation~\ref{sfourdo}; as described in 
detail in (\protect\cite{Unifi} chapters~6 and 9.2, 
\protect\cite{Gener} sections~2.3 and 3.1) and summarised below.}
\label{vallo}
\end{table}

 The initial stage is to write equation~\ref{sfourdo} in the form of 
equation~\ref{lpvno} and to arrange the four $\{v^a\}$ components in a 
$2 \times 2$ Hermitian complex matrix with $\bv_4 \equiv \bh \in \htwc$ 
in a standard way such that
 $L_2(\bv_4)_{\mathrm{SL}(2,\ccc)}= \det(\bh) = 1$ as a matrix 
determinant, where $\sltc$ is the double cover of the Lorentz group 
$\soot$ (\cite{Gener} equations~15 and 18). This structure can then be 
directly subsumed within an augmented  $3 \times 3$ Hermitian complex 
matrix form $L_3(\bv_9)_{\mathrm{SL}(3,\ccc)}= \det(\bv_9) = 1$ for the 
9-component vector $\bv_9 \in \hthc$ with an $\slthc$ symmetry
 (\cite{Gener} equations~16--20).

  The octonions enter at the next stage via the augmentation $\ccc \to 
\ooo$ from the $\slthc$ level to the 27-dimensional form
 $L_3(\bv_{27})_{\mathrm{E}_6\equiv \mathrm{SL}(3,\ooo)}= 
\det(\bv_{27}) = 1$, where $\bv_{27} \in \htho$ is a $3 \times 3$ 
Hermitian octonion matrix (and hence an element of the exceptional 
Jordan algebra), since such a determinant can still be defined for the 
octonion case (\cite{Gener} equation~30--31). In constructing the 
symmetry transformations upon this cubic norm through an octonion 
composition care is needed for the non-associativity property, as 
described explicitly in (\cite{Wang} chapter~4, \cite{Man4} section~6, 
as reviewed in  \cite{Unifi} chapter~6). In particular the 
\mbox{$\det(\bv_{27})$-preserving} symmetry actions are explicitly 
constructed in terms of non-associative $m$-fold \textit{nested} 
$3\times 3$ matrix transformations of the general form
 (\cite{Wang}~subsection~4.1.1, \cite{Unifi} equation~6.30):
\begin{equation}
 \label{nest3o}
    \bv_{27} \: \to \: 
  \mcM_m(\ldots(\mcM_1(\bv_{27})\mcM_1^{\dag})\ldots)\mcM_m^{\dag}
\end{equation}

 The action of the octonion-valued $3\times 3$  matrices $\mcM$ 
(\cite{Unifi} table~6.1 and equations~6.29--6.35) on 
 $\bv_{27} \in \htho$ can be considered a generalisation from  the 
original $\sltc$ double cover of the Lorentz rotations and boosts in 
 \mbox{4-dimensional} spacetime for this higher-dimensional 
augmentation (see also~\cite{Man2} for the intermediate case of 
$\sltwoo$ Lorentz transformations for a 10-dimensional spacetime 
represented by $\htwo$).
 For the  cubic norm on the space $\htho$ it is precisely through the 
octonion non-associativity that there is enough freedom in the 
construction of equation~\ref{nest3o} to identify a 
  full exceptional Lie group $\esi \equiv \sltho$ symmetry structure 
through this compact $3 \times 3$ matrix expression, with some cases of 
$m>1$ nested compositions corresponding to elementary group actions 
(see discussion in~\cite{Unifi} between equations~6.35 and 6.36).

  The final extension in table~\ref{vallo} to 
  $L_4(\bv_{56})_{\mathrm{E}_7}= q(\bv_{56}) = 1$
   is to a 56-dimensional space,
   where the quartic norm $q(\bv_{56})$ on elements of the Freudenthal 
triple system $\bv_{56} \in F(\htho)$ is defined for example 
in~(\cite{Rios} subsection~3.4 and references therein). The space 
$F(\htho)$ can be decomposed as two  
 sets of $\htho$ subcomponents and two further real $\rrr$ 
subcomponents. 
While the quartic norm $q(\bv_{56})$ is not itself a determinant the 
transformations of the $\esi\subset \ese$ subgroup embedding are 
straightforward to describe in acting upon one $\htho$ subspace as 
reviewed above for equation~\ref{nest3o} and the other $\htho$ subspace 
as the complex conjugate representation (as reviewed for \cite{Gener} 
equation~34), while the $\esi \subset \ese$ action is trivial on the 
two real subcomponents. The full set of $\ese$ transformations is 
described in (\cite{Unifi} equations~9.29--9.32 and the accompanying 
references). These $\esi$ and $\ese$ constructions have a known 
interpretation in a context of `generalised spacetimes'  in augmenting 
the Lorentz symmetry of \mbox{4-dimensional} spacetime (\cite{Gunay2} 
equations~5 and 6), while here, as norm-preserving actions upon 
homogeneous polynomial expressions, they are considered as symmetries 
$\hat{G}$ of `generalised proper time' in the form of 
equation~\ref{lpvno}.

   For the present theory it is hence through these constructions of 
the exceptional Lie group symmetries $\esi$ and $\ese$ in 
table~\ref{vallo} that the octonion algebra $\ooo$ enters in a  central 
way in describing forms and symmetries of generalised proper time. This 
is in contrast to the special case of extra spatial dimensions for the 
restricted quadratic form of equation~\ref{sndo} which can be described 
in full using the real algebra $\rrr$ only, expressing this 
augmentation with a full metric $\hat{\eta}$ for $n>4$ and with an 
\mbox{$n$-dimensional} Lorentz symmetry $\sootnm$. It is through the 
further generalisation with $p>2$ in equation~\ref{salphao}, that can 
be written as equation~\ref{lpvno}, that the octonions  enter into the 
explicit mathematical structure, including the unique forms in 
table~\ref{vallo} with the
  symmetries $\hat{G} = \esi \equiv \sltho$ and its extension to 
$\hat{G}=\ese$.

  While we noted near the end of the previous section that the 
one-dimensional continuum of real numbers $\rrr$ can be associated with 
the linear continuum of time, as a feature of essentially all 
successful physical theories since Newton, the idea of a connection 
more generally between division algebras and the concept of time also 
has an historical precedent.  In the 1830s Hamilton argued for a hidden 
meaning of the algebraic properties of the complex numbers $\ccc$ as 
deriving from the `science of pure time'~\cite{Ham1}, based on an 
analysis of the arithmetic properties of two-component elements, or 
`number-couples', associated with two independent finite steps in time. 
Attempts to generalise that construction helped lead Hamilton to the 
discovery of the four-component quaternions $\hhh$~\cite{Ham3}, which 
was soon after followed by the discovery of the octonions $\ooo$ by 
Graves and Cayley in the mid-1840s (as discussed in~\cite{TimeE} 
section~3).

 For the present theory natural mathematical augmentations for 
infinitesimal proper time intervals beyond the 4-dimensional geometric 
spacetime form of equation~\ref{sfourdo} through the generalised 
multi-component form of equations~\ref{salphao} and \ref{lpvno} have 
led directly to the employment of the largest division algebra $\ooo$ 
in describing the symmetries and substructure of these elements of 
proper time.
 Hence, with the octonions here entering through the `structure of 
proper time' rather than the `science of pure time', this theory  
 in a sense sees a return full circle back  to Hamilton's conception of 
the complex numbers and the origins of the complete set of division 
algebras themselves. \mbox{We also note that}
for Hamilton the proposed close connection between algebra and order in 
time complemented and in part was motivated by the
well-established intimate connection between geometry and order in 
physical space  (as quoted in~\cite{TimeE} section~3).

  Here the augmentations for equation~\ref{lpvno}, through to the 
octonion-based vector spaces and symmetries of table~\ref{vallo}, are 
proposed to be key not only to the elementary substructure of time but 
also to the observed empirical structure of matter as related to and 
ordered in the geometric structure of 4-dimensional spacetime. As 
discussed in the previous section the relations between \textit{space}, 
\textit{time} and \textit{matter} are central to any physical theory. 
Here they hinge upon the structure of equation~\ref{sfourpo} through 
which the local geometric forms of 3-dimensional space and 
4-dimensional spacetime are projected out of the full form of proper 
time, with the residual components and the corresponding broken 
symmetry providing the basis for the properties of matter,
 which hence in turn will derive from octonionic structures.
 This symmetry breaking structure has been analysed explicitly through 
each stage of table~\ref{vallo}, with the conclusions regarding the 
derived properties of matter fields summarised in the following 
section.

  While having a simple and natural conceptual basis the bottom line, 
as for any theory, will be whether the present theory actually works in 
terms of accounting for the empirically observed properties of matter, 
in particular as described by the Standard Model as reviewed for 
table~\ref{smgen} in the previous section. The degree to which this 
theory, with its octonion-rich structures, has to date made direct and 
explicit connections with these basic properties of particle physics 
will also be reviewed in the next section.

\section{Symmetry Breaking and the Standard Model}
\label{octo3}

  While the general form of proper time can be \textit{mathematically} 
augmented through the progression of higher-dimensional forms described 
for table~\ref{vallo} the empirical world is still \textit{physically} 
perceived against the background of an extended 4-dimensional spacetime 
manifold $M_4$. The necessary extraction of the substructure 
 $\eta_{ab}\delta x^a \delta x^b$ as a basis for the local Lorentzian 
geometry of $M_4$ from equation~\ref{sfourpo} then necessarily breaks 
the full symmetry $\hat{G}$ of equation~\ref{salphao}.
Correspondingly, subcomponents $\bv_4 \in \TM_4$ are projected onto the 
local 4-dimensional external tangent space out of the full $n$-vector 
$\bv_n$ of equation~\ref{lpvno},  the fragmented components of which 
together with  factors in the breaking of the full symmetry $\hat{G}$  
are interpreted as underlying matter fields in $M_4$
 (see for example~\cite{TimeE} figure~4).
 This symmetry breaking structure has been studied in explicit detail 
through to the case of $\hat{G} = \ese$ with $n=56$ and $p=4$ in 
table~\ref{vallo} with the resulting symmetry breaking pattern for the 
subcomponents of
 $\bv_{56}$ determined (\cite{TimeE} section~4) as summarised here in 
table~\ref{esesm}.
 
\def\rai{+0.2ex}
\begin{table}[htbp]
\centering
\begin{tabular}{|r|ccccc|c|}
 \hline
  \raisebox{-0.5ex}{${56}$} \!\!\!\!\!\!\!
   {\mbox{\raisebox{+0.0ex}{\LARGE{$\diagdown$}}}} \!\!\!\!\!\!
      \mbox{\raisebox{+0.7ex}{$\ese \! \supset\!\!$}} 
	    & \raisebox{\rai}{Lorentz} &
  \raisebox{\rai}{$\!\!\!\times\!\!\!$}
	    & \raisebox{\rai}{$\suth_c$}   &
  \raisebox{\rai}{$\!\!\!\times\!\!\!$} 
		& \raisebox{\rai}{$\uo_Q$} & \raisebox{\rai}{state} \\
 \hline
  $4 \qquad$ & \underline{vector}  & & $\b1$  & & 0 & `$\nu_L$'  \\
	& & & & & &
    \vspace{-18pt} \\ 
	$8 \qquad$ & Dirac & & $\b1$  & & 1 & 
{\LARGE$\binom{\mbox{\normalsize$e_L$}}{\mbox{\normalsize$e_R$}}$} \\ 
   & & & & & &
    \vspace{-15pt} \\  	
   $12 \qquad$ & \underline{scalar} & & $\bbb$ & & $\frac{2}{3}$ & 
   {\LARGE$\binom{\mbox{\normalsize{`$u_L$'}}}
              {\mbox{\normalsize{`$u_R$'}}}$}  \\
      & & & & & &
    \vspace{-15pt} \\
   $24 \qquad$ & Dirac & & $\bbb$ & & $\frac{1}{3}$ & 
{\LARGE$\;\!\binom{\mbox{\normalsize$d_L$}}
                    {\mbox{\normalsize$d_R$}}\!\!\:$} 
	\vspace{-15pt}  \\  & & & & & &								 
					\\  \hline
      & & & & & &
   \vspace{-20pt} 
	\\
     $4 \qquad$ & \underline{vector} & & $\b1$ & & 0 & 
          Higgs ($\bv_4$)
					 \\
      & & & & & &
    \vspace{-22pt} \\
    $4 \qquad$ & scalar & & $\b1$ & & 0 & 
				Yukawa					
   	\vspace{-18pt}  \\  & & & & & &
	  \\
   \hline
  \end{tabular}
  \caption{\setb The symmetry breaking of the $\ese$ quartic form for 
proper time $L_4(\bv_{56})_{\mathrm{E}_7}=1$ and  partitioning of the 
56 components of $\bv_{56} \in F(\htho)$ through the extraction of a 
local Lorentz symmetry acting upon the projected $\bv_4 \in \TM_4$ 
subcomponents. The final column lists the matter field interpretation 
(also \protect\cite{Unifi} equation~9.46, \protect\cite{Gener} 
equation~35--36).}
\label{esesm}
\end{table}   

  As listed in table~\ref{esesm} the derived structures include sets of 
Dirac spinors under the external $\mbox{Lorentz} \subset \ese$ symmetry 
as well as both singlets $\b1$ and triplets $\bbb$ under an internal 
$\suth_c$ symmetry. The latter is interpreted as the colour gauge group 
while the further internal symmetry factor $\uo_Q$, for which the 
fractional relative charge magnitudes listed are fixed by the 
mathematical structure, is interpreted as the gauge group of 
electromagnetism. 
Through direct comparison with table~\ref{smgen} the above properties 
collectively motivate the matter field interpretation as a generation 
of leptons and quarks of the Standard Model as listed in the final 
column of table~\ref{esesm}, albeit with quote marks for the 
\mbox{`$\nu_L$'-lepton} and \mbox{`$u_{L,R}$'-quark} states since the 
corresponding Lorentz structure is incomplete as indicated by the 
underlined entries in the table.
(The distinction between particles and antiparticles, with opposite 
$\uo_Q$ charges, for the full theory will depend upon the dynamics of 
particle states in 4-dimensional spacetime and also upon conventions, 
as for the Standard Model, as discussed towards the end of \cite{Unifi} 
section~8.2).

  Given the very direct nature of the symmetry breaking  the clear 
resemblance of the derived properties in table~\ref{esesm} described 
above to those of
 the Standard Model as listed in
 table~\ref{smgen} is striking.
 In particular the inroads into the properties of the Standard Model 
obtained through generalised proper time go much further 
and in a much more direct manner
 than can be achieved for models with extra spatial dimensions (as 
alluded to in the opening of section~\ref{octo2}),
 that is by models effectively based on equation~\ref{sndo} as a 
restricted case of equations~\ref{salphao} and \ref{lpvno}  with $n>4$ 
but with $p=2$ only. In fact through a similarly direct analysis the 
properties obtained for matter fields in the case of extra spatial 
dimensions bear no resemblance to the Standard Model (as discussed 
for~\cite{Gener} equation~8--9 and figure~1). 
  Moreover the inroads into the Standard Model achieved for the present 
theory continue with a correspondence identified with some of the 
further esoteric features discussed following table~\ref{smgen}, as we 
review here.

 The four components of $\bv_4 \in \TM_4$ projected out of the 
56-dimensional vector \mbox{$\bv_{56} \in F(\htho)$}, as subsumed from 
the components of $\bv_4 \equiv \bh \in \htwc$ for the original 
4-dimensional case in table~\ref{vallo} and as intimately associated 
with the symmetry breaking for the higher-dimensional forms of proper 
time, are associated with a non-standard Higgs in this theory as 
indicated in table~\ref{esesm} and as discussed in (\cite{TimeE} after 
figure~4).
The manner in which the Higgs structure in this theory accounts for the 
`origin of mass' consistently with the mutual relation between 
energy-momentum  and spacetime curvature as conceived in the framework 
of general relativity is described for
  (\cite{Gener} equations~22--24). The spectrum of individual particle 
masses is attributed to Yukawa coupling factors deriving from the 
vacuum values of scalar invariant components (which might also have 
some connection with a `dark sector') as listed in the bottom line of 
table~\ref{esesm} and as proposed in (\cite{Gener} subsection~4.2).

  While a full electroweak symmetry
  $\sutw_L \times \uo_Y$ is not obtained at this stage there are 
elements closely analogous to the electroweak symmetry breaking 
structure
  $\sutw_L \times \uo_Y \to \uo_Q$ identified at the $\esi$ and $\ese$ 
levels, in a manner consistent with the lepton and quark assignments in 
table~\ref{esesm}, which further motivate the association of the 
projected $\bh \equiv \bv_4 \in \TM_4$ components with the Higgs 
(\cite{Unifi} section~8.3). Further, an intrinsic left-right asymmetry, 
a salient feature of weak interactions, is distinctly identified at the 
$\ese$ level (as discussed in \cite{Gener} towards the end of 
subsection~3.1).

  The principal remaining properties of the Standard Model that still 
need to be identified are the correct Lorentz spinor structures for the 
neutrino and $u$-quark states, as underlined in table~\ref{esesm}, 
together with a full electroweak and Higgs sector and a full three 
generations of leptons and quarks, to complete the match with the 
empirical properties
 reviewed for table~\ref{smgen}. These features are all correlated 
through the need to fully identify the weak $\sutw_L$ symmetry 
component, which acts upon doublets of left-handed spinor states while 
mixing across the three generations. 
Here the complete set of properties are predicted to arise through one 
further augmentation beyond the $\ese$ stage of table~\ref{vallo} to  a 
full $\hat{G} = \ee$ symmetry of a proposed octic form for 
equation~\ref{lpvno} that can be written as 
 (\cite{Unifi} section~9.3, \cite{Gener} equation~37):
\begin{equation}
 \label{lvto}
 \lvtfep
\end{equation}

  As well as potentially completing the Standard Model picture through 
an $\ee$ symmetry breaking pattern over a $\bv_4 \in \TM_4$ 
subcomponent projection~\cite{TimeE}, in general terms the possible 
existence of such a 248-dimensional form for proper time can also be 
provisionally connected with prospects for new physics beyond the 
Standard Model. These include the possibility of accommodating two 
right-handed neutrinos $\nu_R$ alongside three generations of $\nu_L$ 
states,
 consistent with the observations of solar and atmospheric neutrino 
oscillations as alluded to in section~\ref{octo1}, and in a manner 
related to the possible implication of a composite Higgs structure (as 
described in detail in \cite{Gener} subsection~4.1). A candidate source 
for dark matter analogous to that of Higgs portal interaction models 
can also be identified as well as further potential for new phenomena 
and tests of the theory (\cite{Gener} subsection~4.2). All of these 
areas of new physics are of significant ongoing interest (see for 
example~\cite{Chang,KrHi,Arca}).

   While the octonions, as uniquely the largest division algebra, are 
central to the progression to the $\hat{G}= \esi$ and $\hat{G} = \ese$ 
 symmetries of generalised proper time in table~\ref{vallo} the 
octonion algebra is also expected to be key to the construction of the 
predicted final augmentation to equation~\ref{lvto}, with $\ee$ 
uniquely the largest exceptional Lie group.
  The property of `octonion triality' is already central to the 
construction of $\esi$ acting on the space $\htho$ through 
equation~\ref{nest3o} which incorporates three overlapping actions of 
$\sltwoo \subset \sltho \equiv \esi$  on three $\htwo \subset \htho$ 
overlapping subspace embeddings (\cite{Wang} chapter~4, \cite{Unifi} 
section~6.4). The construction of the $\ee$ symmetry for 
equation~\ref{lvto}  
  is anticipated to further involve the property of octonion triality 
in an essential way in building upon table~\ref{esesm} to obtain the 
correct Lorentz spinor structure for the neutrino and $u$-quark states, 
and for a full three generations of leptons and quarks, in a direct and 
unambiguous mathematical manner.
  While the alignment of the four stages in table~\ref{vallo} is 
manifest, in terms of a neat series of subspace and subgroup 
embeddings, the need to obtain a full spinor structure and electroweak 
symmetry for the final stage suggests that the outermost `Russian doll' 
for $\ee$ may be facing somewhat \textit{askew} with respect to the 
others (\cite{TimeE} section~5).

 As described in (\cite{TimeE}, \cite{Gener} subsection~3.2)  the quest 
for this $\ee$ action on the octic form of equation~\ref{lvto} is 
seemingly related to a number of other studies, such as regarding the 
construction and interpretation of the $\ee$ entry in the $4 \times 4$ 
\mbox{`magic square'} (see for example~\cite{Evans1}). The further 
development of the present theory might also connect with other 
analyses involving the exceptional Jordan algebra $\htho$~(such as 
\cite{Todo}) and with 
the progress achieved in other physical theories. Indeed 
 through related mathematical structures the octonions are also 
prevalent in the contemporary unification schemes of supersymmetry, 
string theory and M-theory~\cite{Boya,Topp}. In striving to attain 
uniqueness in these or other unified theories a possible means may be 
through a direct and unambiguous link with unique and specific 
mathematical structures such as associated with the octonions and with 
the exceptional Lie groups.    

While abstract constructions with the octonions
 and
 patterns of $\esi$, $\ese$ and $\ee$ symmetry breaking   are known 
\textit{algebraically} to have some correlation with Standard Model 
properties as reviewed in section~\ref{octo1}, here crucially these 
mathematical objects enter through an underlying \textit{conceptual} 
basis for the theory. This theory \textit{originates} from the 
generalisation of proper time beyond the 4-dimensional spacetime form 
of equation~\ref{sfourdo} to equation~\ref{salphao}, as described in 
section~\ref{octo2}, and is \textit{subsequently} found to incorporate 
the octonions and exceptional Lie groups through explicit natural 
mathematical augmentations for equation~\ref{salphao}, written in the 
form of equation~\ref{lpvno} and as described for table~\ref{vallo}.
 \mbox{Finally} a close resemblance with elements of the Standard Model 
is \textit{directly} uncovered as can be seen by comparison of 
table~\ref{smgen} with table~\ref{esesm} and the accompanying 
discussion, and still very much with potential for further progress as 
also discussed in this section regarding the structure of 
equation~\ref{lvto}.      
 Through this theory based upon generalised proper time the octonions 
are found then to be fundamental to understanding the unique properties 
of the Standard Model and physics beyond.

 These observations rely on the norm-preserving octonion transformation 
properties as introduced in equation~\ref{normp} and applied for the 
\mbox{$3 \times 3$} matrix determinant case of equation~\ref{nest3o}, 
with further generalisation presumably needed for the $\ee$ level of 
equation~\ref{lvto}.
 Explicit calculation assimilating the octonion non-associativity 
through the nested bracket structure of equation~\ref{nest3o}
  is needed to determine the correspondence with the full unbroken 
$\esi$ and $\ese$ group properties for table~\ref{vallo}, with a 
related  construction anticipated to also be required for the $\ee$ 
group symmetry stage of equation~\ref{lvto}.

  Although the nested brackets are necessary to construct the full 
symmetry at the $\esi$ and $\ese$ levels, a direct one-to-one 
correspondence can still be identified between the generators of the 
subgroup factors in the top line of table~\ref{esesm} and elements of 
conventional bases employed for these symmetries in the Standard Model. 
In fact the \mbox{$\mbox{Lorentz} \subset \esi \subset \ese$} subgroup 
symmetry is described by six matrix actions $\mcM$ in 
equation~\ref{nest3o} all of which have entries belonging to the 
\textit{same} $\ccc \subset \ooo$ subspace. This means that all 
combinations of actions on individual components of $\bv_{27}\in \htho$ 
are defined within an associative $\hhh \subset \ooo$ subalgebra and 
the brackets in equation~\ref{nest3o} can be removed without ambiguity. 
This allows for a straightforward correspondence with a standard basis 
for Lorentz transformations as described for (\cite{Unifi} 
equations~8.6--8.9).

  On the other hand, as a consequence of octonion triality out the full 
set of 78 independent basis transformations that compose the  full 
$\esi$ symmetry on $\htho$ only the \mbox{14 transformations} of a 
 $\gt \subset \esi$ subgroup are necessarily constructed from nested 
actions (\cite{Man4} section~6, \cite{Unifi} after equation~6.50).  
 The internal symmetry colour gauge group in table~\ref{esesm} is in 
turn obtained as a subgroup $\suth_c \subset \gt$ and hence is also 
necessarily described in terms of nested transformations. However a 
\mbox{one-to-one} correspondence between the generators of this 
$\suth_c$ and a conventional basis of eight Gell-Mann matrices can 
still be identified as described in (\cite{Wang} section~4.7, 
\cite{Unifi} tables~8.4--8.6 with a further note on conventions in the 
three paragraphs following table~6.5).

 As described in section~\ref{octo1} a core feature of 
 the initial interest regarding the possible application of the 
octonions
  in particle physics concerned 
   the $\suth \subset \gt \equiv \mbox{Aut}(\ooo)$ subgroup of the 
octonion automorphism group that might be associated with the $\suth_c$ 
colour gauge symmetry of the Standard Model~\cite{Gunay}.
  In comparison with such models here, as noted above, the $\suth_c$ 
symmetry in table~\ref{esesm} is similarly traced to an $\suth \subset 
\gt \equiv \mbox{Aut}(\ooo)$ symmetry involving transformations upon 
octonion elements that leave invariant the real unit and one imaginary 
unit, with  a link between this $\suth_c$ and that in \cite{Gunay}  
made explicitly in (\cite{Wang} subsection~4.2.3).

 For the present theory based on the general form of proper time in 
equation~\ref{lpvno} a natural full symmetry $\hat{G}$ breaking 
\textit{mechanism} is identified through the necessary distinction of 
the specific substructure of 
 four projected subcomponents  $\bv_4 \in \TM_4$ as a 
  basis for the local geometry of the external 4-dimensional spacetime 
$M_4,$ as described in the opening of this section.
 For the cases in table~\ref{vallo} of the 27 and 56-dimensional forms 
two of the components of $\bv_4 \in \TM_4$ correspond to a real and an 
imaginary unit of an octonion element (with coefficients $a_1$ and 
$a_8$ in equation~\ref{octa}) of $\bv_{27} \in \htho$ and $\bv_{56} \in 
F(\htho)$, 
 as described for (\cite{Unifi} equation~8.4--8.5, \cite{TimeE} 
equation~54).
 The above subgroup
 $\suth_c \subset \gt \subset \esi \subset \ese$ is then interpreted as 
an \textit{internal} 
symmetry precisely since it stabilises these \textit{external} 
components, and the projected $\bv_4 \in \TM_4$ generally, 
 as discussed for  (\cite{Unifi} table~8.3--equation~8.16), and hence 
the symmetry breaking mechanism itself directly \textit{explains} the 
identification of this $\suth_c$ action as an internal gauge symmetry.

 While this octonion-based $\suth_c$ colour gauge symmetry is in itself 
analogous to that employed in~\cite{Gunay,Mori2,Dix1,Fur1}, the 
mathematical structure of the broader connections with the Standard 
Model, deriving from the $\ese$ quartic norm in table~\ref{vallo} as 
described here for table~\ref{esesm}, is unlike that of other schemes 
utilising the division algebras. It is possible however that there 
could be further connections, for example the means of obtaining three 
generations in \cite{Dix2,Fur2} might provide a further guide for the 
uncovering of the full Standard Model through an octonion-based 
construction of the $\ee$ level of equation~\ref{lvto} proposed  here. 
 For the present theory though, rather than being based on abstract 
mathematical structures, 
the initial underlying conceptual motivation as well as the symmetry 
breaking mechanism, through which the connections with physics are 
made, are very different.

 In particular, returning to the question raised towards the end of 
section~\ref{octo1}, concerning \textit{what} the mathematical 
structure of octonions might conceptually represent or act upon to 
connect with the physical world, here it enters through  objects such 
as $\bv_{27} \in \htho$ and $\bv_{56} \in F(\htho)$ in 
table~\ref{vallo}, as also anticipated for the final extension to 
$\bv_{248}$ in equation~\ref{lvto} (\cite{TimeE} equation~74). These 
objects, with the octonions a principal constituent,  represent 
generalised forms of proper time, as concrete realisations of 
equation~\ref{lpvno}.
   Symmetry transformations utilising octonion-valued compositions as 
discussed for equation~\ref{nest3o} then describe the symmetries 
leaving these forms for equation~\ref{lpvno} invariant. This leads to 
the explicit construction of the Lie group symmetries $\esi \equiv 
\sltho$ and $\ese$, with the predicted extension to an $\ee$-related 
action for equation~\ref{lvto} also proposed to incorporate properties 
of the octonions in essential way.

  Through the symmetry breaking mechanism, analysed through to the 
level of the $\ese$ quartic form of proper time in table~\ref{vallo}, 
lepton, quark and Higgs states are associated with the fragmented 
components of the octonion-based $\bv_{56} \in F(\htho)$ structure, 
while external Lorentz and internal strong and electromagnetic gauge 
groups are identified in the breaking of the octonion-based full 
symmetry $\hat{G} = \ese$ action on this form as summarised in 
table~\ref{esesm}.
  A correspondence can then be made between 
features of this symmetry breaking structure and a generation of 
leptons and quarks, as well as other aspects of the Standard Model 
through comparison with table~\ref{smgen} and the accompanying 
paragraphs, as described in this section.
	Hence the present theory profusely utilises the octonion algebra 
but in a manner that is firmly grounded in a clear conceptual picture,  
incorporating both the structure and symmetry of generalised proper 
time, while also directly connecting the octonions with the esoteric 
structures of particle physics.     

\section{Summary and Outlook}
\label{octo4}

  In light of the interest and debate concerning the possibility of a 
fundamental role for the octonions in particle physics (see for example 
\cite{Gunay73,Gunay,SoLo,Mori2,Dix1,Dix2,Mori15,Fur1,Fur2,Rowl}) in 
this paper we have described how such a connection can be achieved in 
an explicit, unambiguous and predictive manner through the structure of 
generalised proper time. Here we summarise the main points:

\begin{itemize}

\item In augmenting beyond the local 4-dimensional spacetime structure 
for a proper time interval in equation~\ref{sfourdo} it is possible to 
generalise beyond the case of extra spatial dimensions in 
equation~\ref{sndo} to the form of equation~\ref{salphao} with $n>4$ 
and also $p>2$, dropping the assumption of a quadratic $p=2$ form.

\item With the theory interpreted as a simple generalisation of a 
proper time interval, as can be expressed in the form of 
equation~\ref{lpvno}, the octonions are found to be extensively 
employed in constructing a natural and unique mathematical progression 
to higher-dimensional forms as described for table~\ref{vallo}.

\item Matter fields deriving directly from the necessary symmetry 
breaking over a 4-dimensional spacetime background exhibit a 
non-trivial correspondence with properties of the Standard Model, based 
on comparison of tables~\ref{smgen} and \ref{esesm}, affording the 
octonions a direct application in particle physics.  
 
\end{itemize}

  The theory hence provides a tangible conceptual link between the 
octonions and particle physics, without relying primarily on a 
potentially arbitrary or subjective notion of mathematical beauty or 
uniqueness as the chief heuristic guide. The direct and incisive 
foothold achieved within the esoteric structures of the Standard Model 
 includes the identification of Dirac spinors, $\suth_c$ singlets and 
triplets, $\uo_Q$ fractional charges and an intrinsic left-right 
asymmetry, as well as other features of electroweak and Higgs physics, 
as discussed for table~\ref{esesm}. These empirical connections go much 
further than can be obtained directly with models restricted to extra 
spatial dimensions as discussed in the opening of section~\ref{octo2}.

 Seen as a generalisation from models with extra spatial dimensions, 
the theory has historical roots  dating back to the unified field 
theories of circa the 1920s (\cite{Gener} subsections~1.2 and 5.1).
  The theory also has significant historical  links on the mathematical 
side dating back to the origins of the division algebras themselves. 
From the continuum of real numbers $\rrr$ representing the linear 
`continuous flow of time', as alluded to in section~\ref{octo1} and as 
utilised since Newton, through Hamilton's interpretation of the complex 
numbers $\ccc$ as an `algebra of pure time' in the 1830s, leading to 
his discovery of the quaternions $\hhh$ as reviewed in 
section~\ref{octo2}, here the octonions $\ooo$ serve a pivotal function 
in `generalised forms of proper time'. 

 This connection with \textit{time}, as arguably the most fundamental 
entity in physics, is key to establishing the link between the rich 
structures of the octonions and the rich structures of matter in 
4-dimensional spacetime as described by the Standard Model.  
While the octonion algebra in itself is too simple to encompass the 
full range of Standard Model features, leading to ambiguity in how to 
proceed in building models based upon the abstract properties of the 
division algebras  as reviewed in section~\ref{octo1}, the concept of 
generalised proper time provides a well-defined means of accommodating 
octonions in a broader yet unique framework for connecting with 
particle physics. 
  Further historical and philosophical background to the conceptual 
picture is presented in~\cite{Struct}.

 The future outlook concerns the prospects for the further development 
of the theory beyond the forms for proper time in table~\ref{vallo} to 
the predicted $\ee$ action on an octic form in equation~\ref{lvto}, the 
construction of which is anticipated to inherently incorporate and 
utilise properties such as octonion triality.  
  While relatively recent literature has provided an off-the-shelf 
guide to the $\esi$ and $\ese$ levels of the theory (for 
example~\cite{Wang,Rios}), allowing the explicit calculations 
underlying table~\ref{esesm}, further mathematical understanding for 
the $\ee$ level, including for example in connection with~\cite{Gunay2} 
and in relation to the magic square~\cite{Evans1,Todo},
  will be needed in parallel with the further development of this 
theory as discussed in section~\ref{octo3}. 

 Such further development will be required in the attempt to both 
complete the match with the Standard Model and to make explicit 
predictions beyond for new physics as also noted in 
section~\ref{octo3}.
  However, given the simplicity of the underlying conceptual picture 
and the non-trivial direct connections with the Standard Model already 
attained through explicit calculations, the theory provides a very firm 
basis for a rudimentary and intrinsic role for the octonions in 
elucidating the origins of the elementary structures of particle 
physics.


{\small
{\setlength{\baselineskip}{0.63\baselineskip}

\par}
}


\par}

\end{document}